\documentclass[final,5p,times,twocolumn]{elsarticle}
\usepackage{amssymb}
\usepackage{amsmath}
\usepackage{graphicx}

\makeatletter
\def\@author#1{\g@addto@macro\elsauthors{\normalsize%
    \def\baselinestretch{1}%
    \upshape\authorsep#1\unskip\textsuperscript{%
      \ifx\@fnmark\@empty\else\unskip\sep\@fnmark\let\sep=,\fi
      \ifx\@corref\@empty\else\unskip\sep\@corref\let\sep=,\fi
      }%
    \def\authorsep{\unskip,\space}%
    \global\let\@fnmark\@empty
    \global\let\@corref\@empty  
    \global\let\sep\@empty}%
    \@eadauthor={#1}
}
\makeatother

\journal{Nuclear Instruments and Methods in Research A}

\newcommand{\degrees}{\ensuremath{^\circ}}
\begin{document}
\begin{frontmatter}

\author{J.D. Maxwell\corref{cor1}}
\ead{jdmax@mit.edu}
\author{C.S. Epstein}
\author{R.G. Milner}

\address{Laboratory for Nuclear Science, Massachusetts Institute of Technology, Cambridge, MA 02139 USA}
\cortext[cor1]{Corresponding author}

\title{Diffusive Transfer of Polarized $^3$He Gas through Depolarizing Magnetic Gradients}

\begin{abstract}
Transfer of polarized $^3$He gas across spatially varying magnetic fields will facilitate a new source of polarized $^3$He ions for particle accelerators.  
In this context, depolarization of atoms as they pass through regions of significant transverse field gradients is a major concern.  To understand these depolarization effects, we have built a system consisting of a Helmholtz coil pair and a solenoid, both with central magnetic fields of order 30 gauss.  The atoms are polarized via metastability exchange optical pumping in the Helmholtz coil and are in diffusive contact via a glass tube with a second test cell in the solenoid. We have carried out measurements of the spin relaxation during transfer of polarization in $^3$He at 1\,torr  by diffusion. We explore the use of measurements of the loss of polarization taken in one cell to infer the polarization in the other cell.
\end{abstract}


\end{frontmatter}


\section{Introduction}

The addition of a polarized $^3$He ion source to the relativistic heavy ion collider (RHIC) at Brookhaven National Laboratory (BNL) would enable a new program of measurements with an effective polarized neutron beam, the merits of which become particularly attractive with a future electron--ion collider~\cite{deshpande}.  A proposal to realize such a polarized neutron source~\cite{zelenski} leverages the electron beam ion source (EBIS) currently in use at BNL~\cite{alessi}. Helium-3 gas would be polarized via metastability exchange optical pumping and introduced into EBIS for ionization and extraction. 
A key challenge in the technique is the transfer of polarized gas from a polarizer in a uniform magnetic field, through a stray field of potentially depolarizing magnetic field gradients, into the EBIS 5\,T solenoid field.

Metastability exchange optical pumping (MEOP) allows enhanced polarization of $^3$He nuclei in a low pressure gas using a uniform magnetic field and circularly polarized light \cite{colgrove}. An RF discharge in the gas promotes a small fraction of the atoms into the 2$^3$S$_1$ metastable state. Transitions from the 2$^3$S$_1$ into the 2$^3$P$_0$ states are driven using 1083\,nm laser light, which will change the magnetic quantum number by $\pm 1$ depending on the circular polarization of the light. The polarization thus induced in the metastable population is then transferred into the ground state population via metastability exchange collisions. 

The planned source would produce polarized $^3$He near 1\,torr, where low field MEOP of $^3$He proceeds optimally, before transferring the gas into EBIS, which operates at just $10^{-7}$\,torr. After ionization and extraction, EBIS would deliver approximately $10^{11}$ ions for each 20\,$\mu$sec pulse. For each pulse, a small sample of the gas in the 1\,torr polarizer cell would be released into the transfer line via a quick-pulse valve, so that the EBIS operating pressure of $10^{-7}$\,torr is maintained. Although this means that gas following the transfer line into EBIS will be far below 1\,torr, we can study the effects of the depolarizing magnetic gradients traversed by the path by observing polarization relaxation in a gas following the same path at 1\,torr. 
In this work, the transfer of polarized gas by diffusion at 1\,torr between two regions of uniform magnetic field was studied in a smaller scale in preparation for tests in the EBIS stray field.

\subsection{Depolarization in Field Gradients}

In a gas of atoms spin-aligned to a holding field $B_0$, Brownian motion of the atoms through a region of space containing magnetic field gradients $\Delta B_t$ transverse to this holding field will cause the ensemble polarization to relax with time constant $\tau$, as shown by Schearer and Walters \cite{schearer}:
\begin{equation}
\label{eq:relax}
\frac{1}{\tau} = \frac{2}{3}\frac{\vert \Delta B_t\vert ^2}{\vert B_0 \vert ^2}\langle v^2 \rangle \frac{\tau_c}{\omega_0^2 \tau_c^2 +1},
\end{equation}   
for particle velocity $v$, mean time between collisions $\tau_c$, and Larmor frequency $\omega_0$ of the nucleus in the holding field. The particle velocity and mean time between collisions are apparent from the temperature and pressure of the gas, and the Larmor frequency is calculable from the holding field and gyromagnetic ratio of the spin, $\omega_0=\gamma B_0 $.  The magnitude of the transverse gradient $\vert \Delta B_t\vert^2$ is given by $(\partial B_t/\partial x)^2 + (\partial B_t/\partial y)^2 + (\partial B_t/\partial z)^2$, taking into account the time-dependent transverse fields the particle experiences while undergoing Brownian motion in three dimensions.  For a field with axial symmetry, $\vert \Delta B_t\vert^2= (\partial B_\rho/\partial \rho)^2 + (\partial B_\rho/\partial z)^2$.


To maintain the polarization of the gas, and thus extend the relaxation time, simply minimizing the ratio of transverse gradients to the holding field is desirable. If we instead consider a constant holding field and transverse field gradients, a minimum relaxation time $\tau$ results when $\tau_c=\omega_0$. Thus spin dis-alignment is most likely to occur when the vibration frequency of the atom due to Brownian motion matches its precession frequency in the magnetic holding field. 

\section{Apparatus}

To observe the effects of depolarizing gradients, Helium-3 gas was polarized in a pumping cell and was passed through a region of sizable transverse field gradients before arriving in a test cell.  Measurements of the nuclear polarization in both cells were made by observing the optical polarization of 667 nm discharge light.  The pumping and test cells were connected by a glass transfer tube with an all-glass valve near the pumping cell to control gas flow.  To ensure gas purity in the cells, the system was baked under vacuum before $^3$He was introduced, and after filling to near 1\,torr, an SAES ST172 getter pump actively removed contaminants which leaked into the volume.

\subsection{Polarizer}

\begin{figure}
\begin{center}
\includegraphics[width=3in]{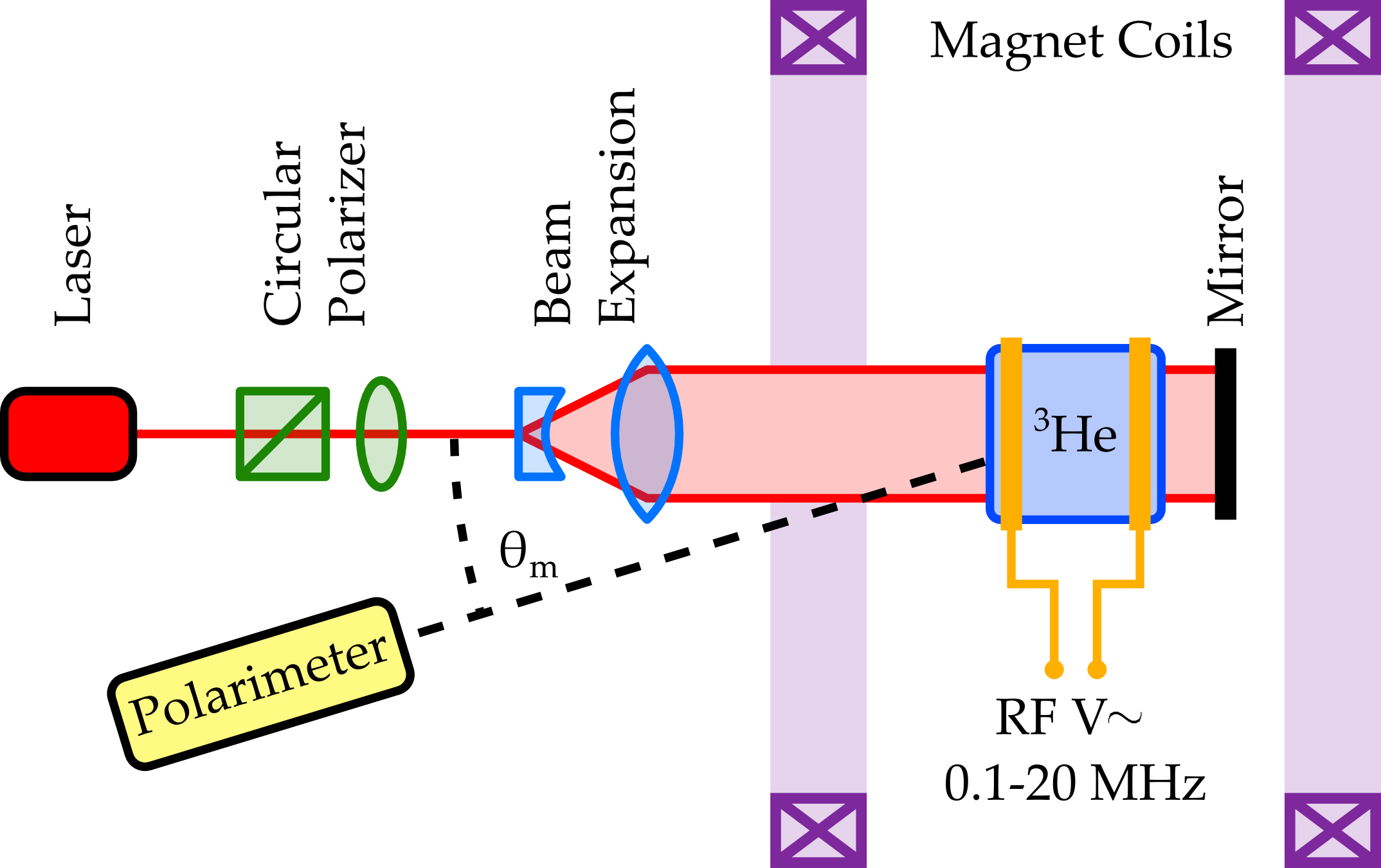}
\end{center}
\caption{Diagram of $^3$He polarizer and polarimeter setup.}
\label{fig:setup}
\end{figure}

Figure \ref{fig:setup} shows a simplified diagram of our MEOP polarizing apparatus.  A weak radio frequency (RF) discharge between 0.1 and 20\,MHz produces a population of metastable atoms in the borosilicate glass pumping cell, which is in a uniform 30\,G magnetic holding field produced by 30\,cm Helmholtz coils.  1083\,nm laser light is circularly polarized and expanded to illuminate the volume of the cell, and a dielectric mirror reflects the laser light back through the cell a second time to increase pumping efficiency.

By observing the circular polarization of the 668\,nm light emitted in 3$^1$D$_2$ to 2$^1$P$_1$ transitions in the discharge \cite{pavlovic}, we obtain a measure of the electron polarization, which is related to the $^3$He nuclear polarization by a correction factor. This correction is heavily dependent on the pressure of the gas, and was determined by NMR calibrations in reference \cite{lorenzon}; for example, the ratio of $^3$He nuclear polarization to electron polarization is 8.37 at 1.010\,torr. Our polarimeter, based on a liquid crystal variable wave plate and described in detail in reference \cite{maxwell}, measures the circular polarization of the 668\,nm discharge light in the gas cell at an angle $\theta_m$. A factor of $1/\cos\theta_m$ corrects for this angular offset.

The 1083\,nm pumping light is provided by a Keopsys ytterbium fiber laser, and is delivered to the polarizer through a polarization-maintaining optical fiber.  The light emerges from the fiber head linearly-polarized, but a polarizing beam-splitting cube is used to ensure full linear polarization.  The light is then circularly polarized using a zero-order $\lambda/4$ wave plate at 45\degrees. To locate the C$_8$ and C$_9$ absorption peaks which correspond to the 2$^3$S$_1$ to 2$^3$P$_0$ transitions of interest, the laser wavelength is swept using a 0--10\,V ramp input control. At resonance, the laser light is maximally absorbed and re-radiated at 90\degrees~to the beam axis; a photodiode next to the pumping cell measures this re-radiated light.  As the offset voltage is swept, the response of this photodiode is mapped and the peaks fit to a Gaussian to return the offset voltage corresponding to each resonance. 

\subsection{Secondary Field and Test Cell}

\begin{figure}
\begin{center}
\includegraphics[width=3.2in]{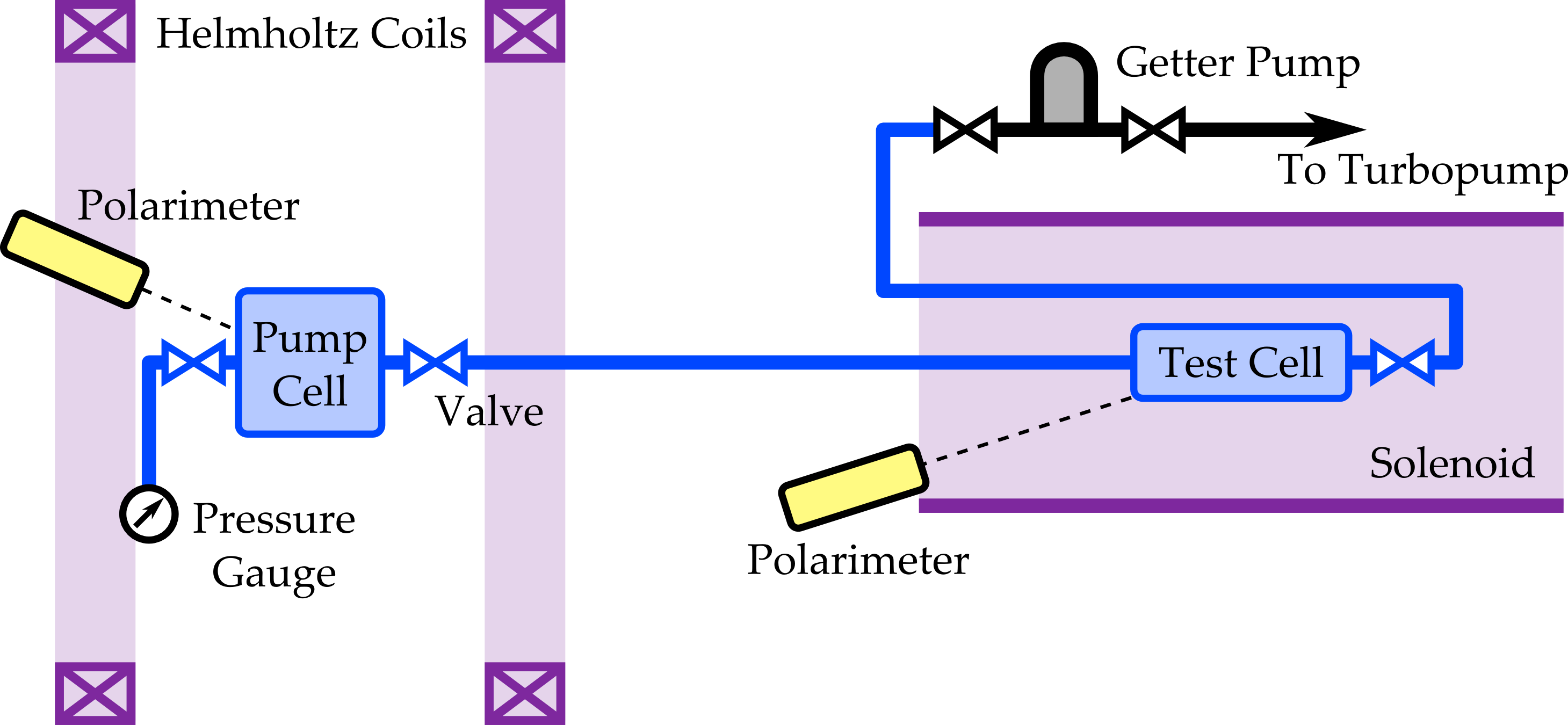}
\end{center}
\caption{Diagram of two-cell polarized gas transfer apparatus. $^3$He gas is polarized in the pumping cell, and can be put into diffusive contact with the test cell by opening a valve.}
\label{fig:flow}
\end{figure}

The test cell is of different dimensions but similar volume to the pumping cell, and is held in a 30\,G solenoidal magnetic field. Figure \ref{fig:flow} shows the layout of the two cells and supporting plumbing. One all-glass valve with viton O-rings was located close to the pumping cell to control the communication of gas between the cells, while secondary valves separated pressure gauges and pumps from the gas volumes during testing.

\begin{figure}
\begin{center}
\includegraphics[width=3.5in]{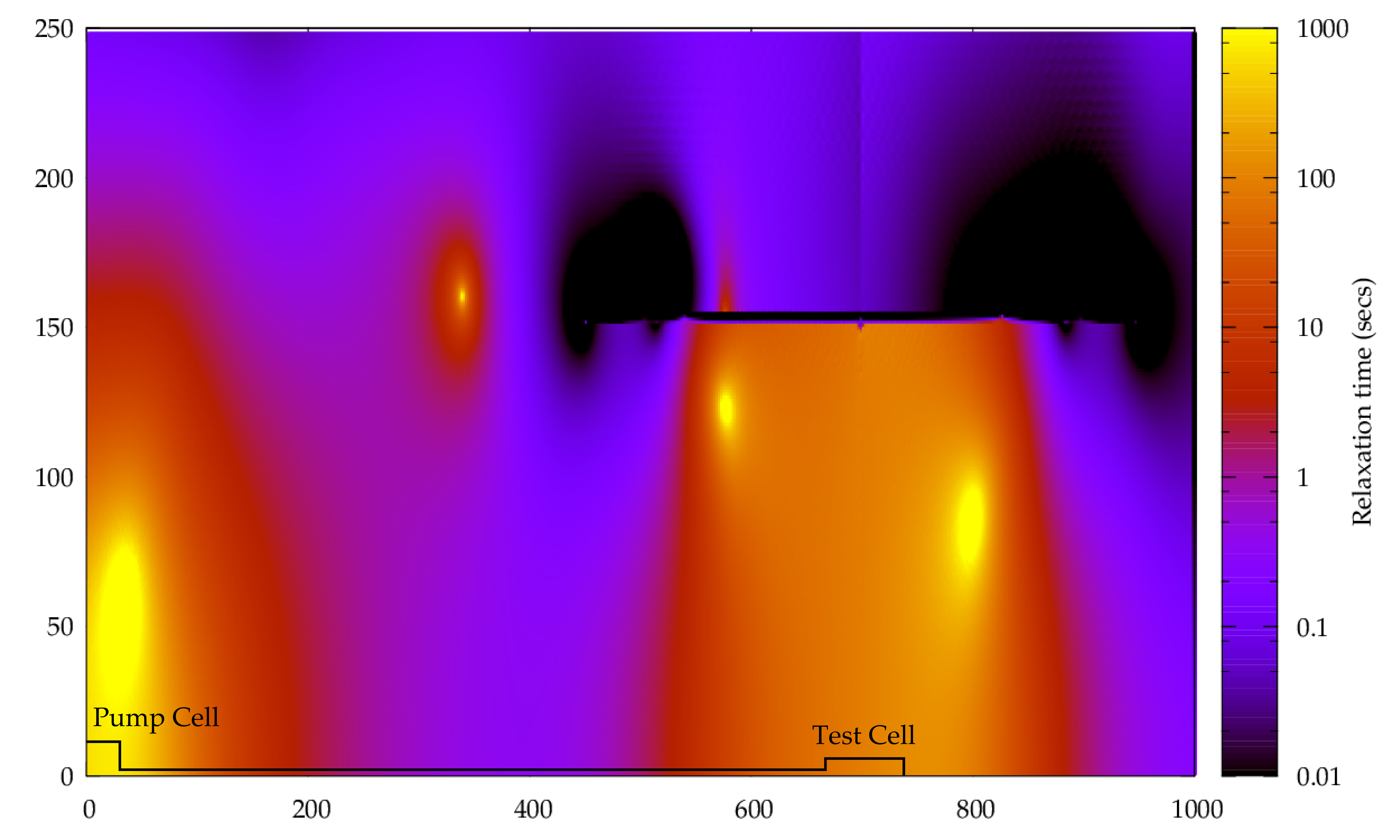}
\end{center}
\caption{Map in mm of expected relaxation time due to transverse field gradients in the two-cell setup. Approximate locations of cells and transfer line are overlaid, with the pumping cell centered at (0,0)\,mm in the center of the Helmholtz field, and test cell centered at (700,0)\,mm, in the solenoid field.}
\label{fig:depol}
\end{figure}

Using equation \ref{eq:relax} and an approximate field map of the area in and around our glass cells, we can produce an expected relaxation time for each point in space. Figure \ref{fig:depol} shows the relaxation time map for our two magnets, with (0,0)\,mm corresponding to the approximate location of the pumping cell, and (700,0)\,mm corresponding to the test cell. Brighter spots inside the Helmholtz pair and solenoid result from uniform fields, and darker spots represent areas where bending field lines lead to fast relaxation.

The transfer path between the two cells roughly follows the $x$-axis from (0,0) to (700,0). We can form a mean relaxation time for this path by taking an average of the inverse relaxation times at each point in the path.  In this configuration, the average relaxation time for 1\,torr helium following the path from the center of the Helmholtz coils to the center of the solenoid is calculated from the field map as 1.2 seconds. Comparing this time to a similar estimate for transfer into the EBIS 5\,T solenoid in the final source design, 1.4 seconds, we find this small scale test to be encouragingly analogous.

\section{Diffusive Transfer between Cells}

\begin{figure}
\begin{center}
\includegraphics[width=3.5in]{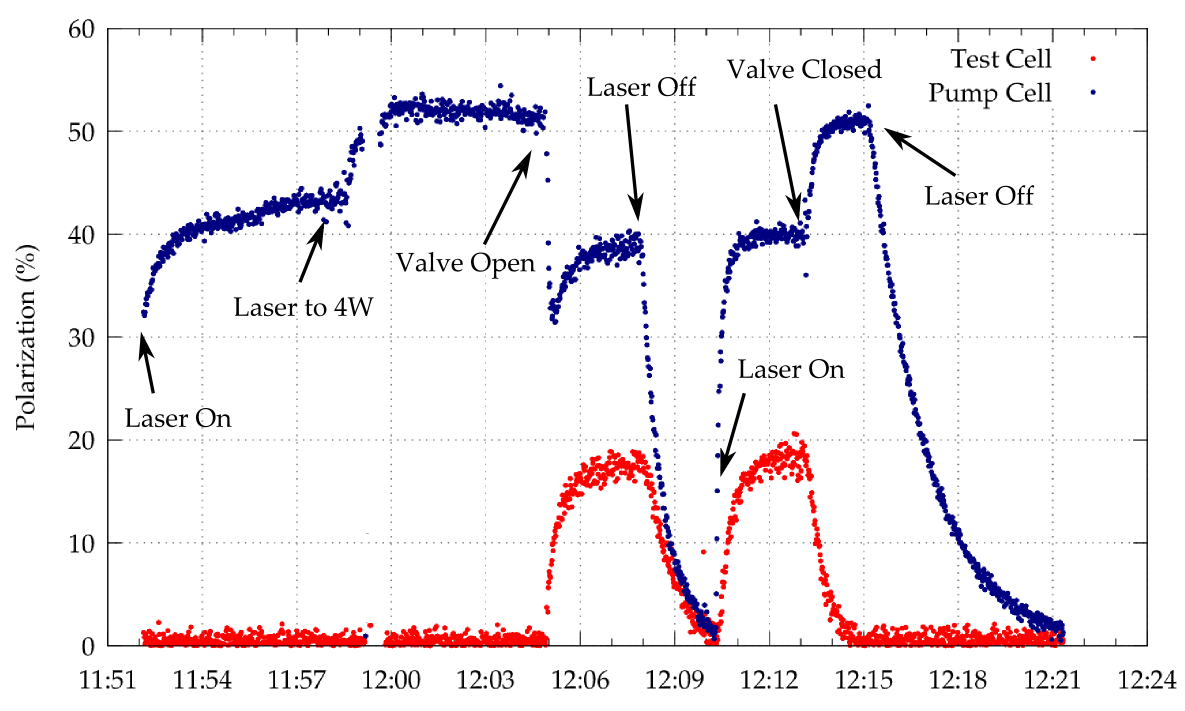}
\end{center}
\caption{Polarization in pump cell (upper points) and test cell (lower points) over time during test of polarization diffusion between cells.}
\label{fig:pol}
\end{figure}

Simultaneous measurements of the polarization in the pumping and test cells were made using polarimeters directed at each. Figure \ref{fig:pol} shows the polarization over time in both cells during pumping, relaxation and transfer. To start, the pumping cell is closed off from the rest of the system. As the pumping laser is turned on and increased to 4\,W, the polarization in the pumping cell (shown in blue) increases to a steady value near 52\%. When the valve separating the 2 cells is opened at approximately 12:04, the polarized gas in the pumping cell begins to mix with the unpolarized gas in the test cell, causing a quick drop in the pumping cell and rise in the test cell. As optical pumping acts on the ensemble volume, the polarization in the pumping and test cells rise to steady values near 38\% and 18\%, respectively. At 12:08, the pumping laser is turned off, allowing both cells to relax to zero, before the laser is returned and each cell returns to its steady polarization. At 12:13, the valve connecting the cells is closed, so the laser pumps the pumping cell polarization back to 52\% and the test cell relaxes to zero without the influx of polarized gas. Finally, the laser is again turned off, and the pumping cell polarization relaxes to zero with a relaxation time of 109 seconds.

These data show that despite the region of depolarizing field gradients between the cells, the diffusive contact between them overcomes this depolarization to allow net polarization in the test cell. The same test performed with the test cell's 30\,G solenoid off results in no polarization in the test cell, as the holding field goes to zero and depolarizing gradients dominate.


\subsection{Modeling Polarization Relaxation in Two Cells}

The interplay between the depolarizing influences in each cell and the diffusion between them can be discerned by studying the relaxation behavior when the pumping laser is shuttered.
Following similar work in polarized helium target cells \cite{jones}, a simplified differential model of the relaxation in each cell can be expressed in terms of the nuclear polarization of the gas in the pumping and test cells $M_p$ and $M_t$, the number of atoms in each cell $N_p$ and $N_t$, the relaxation of the gas in each cell $\tau_p$ and $\tau_t$ and the exchange time of the gas between the cells due to diffusion $t_{\textrm{ex}}$:

\begin{equation}
\label{eq:model}
\begin{pmatrix}
\dot{M}_p(t)\\
\dot{M}_t(t)
\end{pmatrix}
= 
\begin{pmatrix}
-\left(\frac{1}{\tau_p} + \frac{N_t}{N}\frac{1}{t_{\textrm{ex}}}\right) & \frac{N_t}{N}\frac{1}{t_{\textrm{ex}}}\\
\frac{N_p}{N}\frac{1}{t_{\textrm{ex}}} & -\left(\frac{1}{\tau_t} + \frac{N_p}{N}\frac{1}{t_{\textrm{ex}}}\right)
\end{pmatrix}
\begin{pmatrix}
{M}_p(t)\\
{M}_t(t)
\end{pmatrix}.
\end{equation}

The solution is a sum of exponentials:
\begin{equation}
\label{eq:exp}
\begin{split}
M_p(t) &= a_se^{-t/\tau_s}+a_le^{-t/\tau_l} \\
M_t(t) &= a_{st}e^{-t/\tau_s}+a_{lt}e^{-t/\tau_l}
\end{split}
\end{equation}
where new decay constants $\tau_l$ and $\tau_s$ are measurable from fits of exponential decays of the polarization during relaxation. These fit constants are given in terms of the time constants of the system as:
\begin{equation}
\label{eq:const}
\begin{split}
\frac{1}{\tau_{s,l}} &= \frac{1}{1}\left( \frac{1}{\tau_p} +\frac{1}{\tau_t}+\frac{1}{t_{\textrm{ex}}}\right) \\&\pm \frac{1}{2}\sqrt{\frac{1}{t_{\textrm{ex}}^2}+ \left( \frac{1}{\tau_p} -\frac{1}{\tau_t}\right)^2  + \frac{2}{t_{\textrm{ex}}} \left( \frac{N_t-N_p}{N}\right)\left( \frac{1}{\tau_p} -\frac{1}{\tau_t}\right) }.
\end{split}
\end{equation}
The fit amplitudes $a_s$, $a_l$, $a_{st}$ and $a_{lt}$ can also be expressed in terms of the populations, time constants and initial polarizations of each cell. While we refer the reader to reference \cite{jones} for the full expressions, it should be noted that the amplitude fit constants, $a_s$ and $a_l$ sum to give the initial pump cell polarization and $a_{st}$ and $a_{lt}$ subtract to give the initial test cell polarization. Although the two time constants $\tau_l$ and $\tau_s$ are the same for the pumping cell and test cell expressions in equation \ref{eq:exp}, the four amplitude constants are distinct.

In these expressions we treat the transfer line, whose volume is an order of magnitude less than either cell, as part of the test cell. As the constriction of the glass valve means transfer between the line and the pumping cell is less than that between the line and the test cell, this should be a reasonable estimate, though potentially misleading. In reference \cite{jones}, the two cells were connected by a shorter, larger diameter line, and the relaxation in the system occurred primarily within each cell, either due to discharge or the incident electron beam. In our case, a significant depolarizing influence occurs in the transfer line, which will throw off the model.

\begin{figure}
\begin{center}
\includegraphics[width=3.3in]{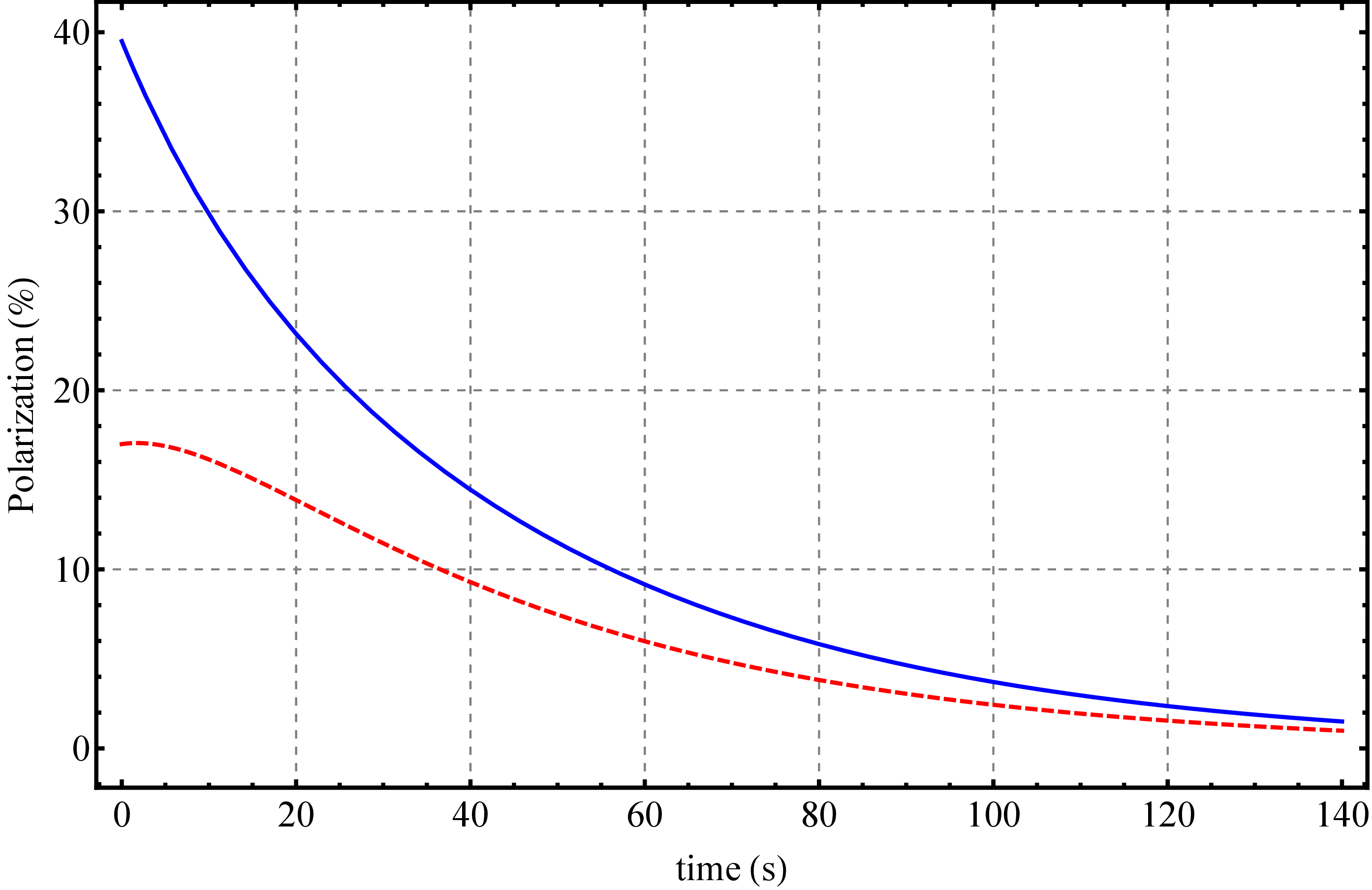}
\end{center}
\caption{Model results for the polarization in the pumping (blue) and test (red) cells during relaxation.}
\label{fig:relax_rates}
\end{figure}

Using estimates of the relaxation in the pumping cell and gas exchange time, we can use this model to plot the relaxation of both cells from a given starting pump cell polarization. Figure \ref{fig:relax_rates} shows the modeled relaxation using a pump cell relaxation time of 90 seconds and exchange time of 15 seconds---values taken from measurements of relaxation in the pumping cell alone and a simple estimate of diffusion. With an eye on the relaxation measurements from figure \ref{fig:pol}, we choose starting polarization of 40\% and 18\% for each cell, and a test cell relaxation of 30 seconds. Both resulting curves shown in figure \ref{fig:relax_rates} have exponential time constants of 46 and 11 seconds, the short constant tending to bring the curves together, and the long constant tending to bring them both to zero. 

Note that in reference \cite{jones}, the relaxation rates $\tau_p$ and $\tau_t$ are much longer than the exchange time $t_{\textrm{ex}}$. This constraint means the fit time constants $\tau_l$ and $\tau_s$ approximate the average relaxation for the ensemble volume and the exchange time, respectively. In our case, the relaxation rates and transfer rate are much closer together, so that the resulting fit constants contain significant mixing terms from all input time constants.

\begin{figure}
\begin{center}
\includegraphics[width=3.5in]{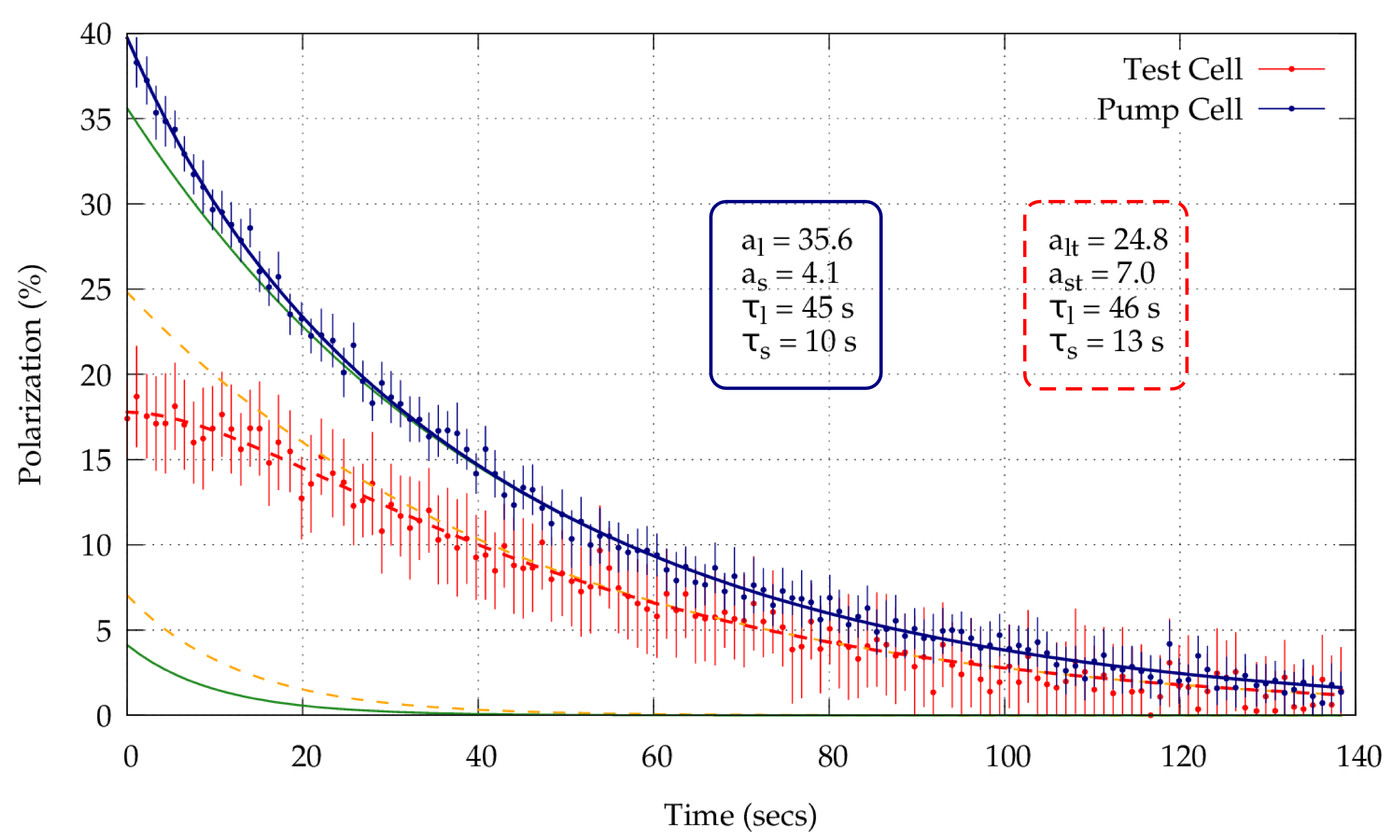}
\end{center}
\caption{Relaxation measurements for pump (blue) and test (red) cells, showing sum of exponential fits to both as solid and dashed lines, respectively. The two exponential curves for each fit which add or subtract to form the respective fit curves are shown in green for the pump cell, and orange for the test cell. The four fit constants for each curve are shown in correspondingly colored boxes.}
\label{fig:laser_off}
\end{figure}

We can now bring this model to bear on the relaxation of both cells from figure \ref{fig:pol}, beginning at around 12:08. Figure \ref{fig:laser_off} shows these relaxation data, as well as exponential fits following equation \ref{eq:exp}. In blue is the sum of exponential fits to the pumping cell curve, with the two exponentials in the sum shown as green curves. Likewise, the red curve shows the test cell fit, with the two exponentials in the difference in orange. 

By design, figures \ref{fig:relax_rates} and \ref{fig:laser_off} are in close agreement, but with caveats.  The long (46~s and 45~s) decay constants in figure \ref{fig:laser_off} agree to within the 10\% fit error, while the short decay constants (13~s and 10~s) disagree slightly. 
This is a symptom of inaccuracy in our model, primarily from the supposition that the transfer line can be treated as a part of the test cell and distinct from the pump cell when significant relaxation occurs in the volume of the transfer line.


\section{Inferring Polarization via Diffusion}

Accepting a degree of inaccuracy from the model, we see that measuring the polarization in both cells provides redundant information. The fit constants $a_{st}$ and $a_{lt}$, extracted from the test cell fit, can be determined using the fit constants from the pumping cell fit and an estimate of the pumping cell relaxation time. In this way, the initial polarization in the test cell can be inferred using only relaxation data from the pumping cell. 

Applying this technique to the data from figure \ref{fig:laser_off}, we supply the model the four pumping cell fit constants and a pumping cell relaxation time of 109 seconds. This results in an exchange time of 15 seconds, a test cell relaxation time of 25 seconds and a starting polarization of 16.8\% in the test cell. This underestimates slightly the measured value for the starting test cell polarization using the second polarimeter, 17.8\%.

\begin{figure}
\begin{center}
\includegraphics[width=3.5in]{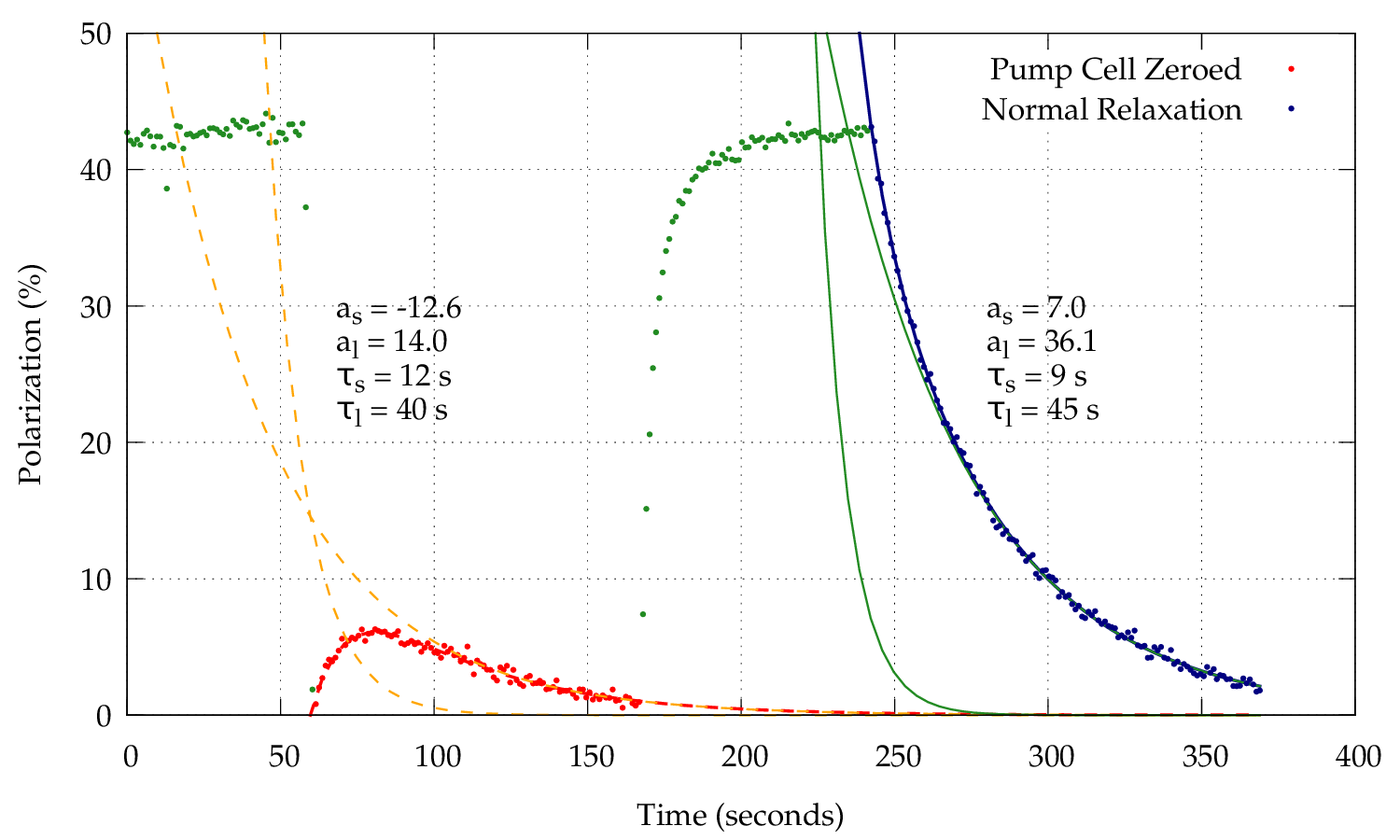}
\end{center}
\caption{Polarization and relaxation over time measured in only the pumping cell while in diffusive contact with the test cell. At about 50 seconds, the pumping laser is shuttered and polarization in the pumping cell zeroed using a moving magnet. In dashed red, the cells mix and relax to zero. At 250 seconds, the pumping laser is shuttered and the cells relax from maximum in solid blue. Both relaxations are fit with a sum of exponentials, and the exponentials in each sum are shown in dashed orange and solid green.}
\label{fig:kill_pol}
\end{figure}

Figure \ref{fig:kill_pol} shows another set of data, taken to further investigate this inference method. Here we measure only the polarization in the pumping cell, which is in diffusive contact with the test cell over a time in which we make two relaxation rate tests. In this case, the measurement is taken in the pump cell with the discharge on, while the test cell discharge remains off, removing the depolarization contribution due to discharge in the test cell. At roughly 50 seconds, the pumping laser is shuttered and a moving magnet is used to quickly depolarize the pumping cell. The polarization in the pumping cell rises as polarized gas from the test cell diffuses back, before dropping again from depolarizing effects. The pumping laser is then re-applied to return the pumping cell to high polarization, before being shuttered at about 250 seconds to allow both cells to relax from their maximal polarizations. These two cases should be essentially identical in our model, with the only difference being the starting polarization of the pumping cell.

The fit constants shown in figure \ref{fig:kill_pol} can be used to calculate the initial polarization in the test cell. We again notice that the time constants for these fits do not match, particularly the short constants, 12 and 9 seconds, which differ by more than the fit error of about 10\%. To infer the initial polarization in the test cell, these fits were used, along with a measured pumping cell relaxation of 126\,s. The fit to the dashed red curve of the zeroed initial polarization decay gives a starting test cell polarization of 32\%, while the decay in solid blue give a starting test cell polarization of 15\%. As a direct measurement of the polarization in the test cell under similar conditions resulted in a polarization of 24\%, these inferred results over and underestimate the true polarization, respectively.

The discrepancy between the inferred polarization in the test cell from the two cases with differing initial pump cell polarizations is problematic, but interesting. 
We speculate that inaccuracy in the model, which causes an underestimation of the inferred polarization in the case of both cells relaxing from high polarization, also creates the overestimation when the pumping cell starts at zero with the test cell polarized. The model attributes any relaxation occurring in the transfer line to the test cell, but the effective relaxation time in the pumping cell is certainly reduced from that when it is closed. In the case of both cells decaying from high polarization, the model responds to an overestimated pumping cell relaxation time by attributing faster relaxation time, and lower initial polarization pulling down the value in the pumping cell, to the test cell. In the case where the pumping cell is zeroed and the polarized gas returns from the test cell, the model again attributes too much relaxation to the test cell, which now requires a higher initial test cell polarization to account for the polarized gas which arrives at the pumping cell.

To illustrate this point, we can simply provide the model with a lower pumping cell relaxation time. With a pump cell relaxation time of 70 seconds rather than the 126 seconds which had been measured with the cell closed, the model returns much more consistent results. For the initially zero pumping cell case, the initial test cell polarization is now 26\%, while for the decay from high polarization case, the initial test cell is instead 23\%. We can see the estimates now converging on a more realistic result, with the mid-point between the two cases canceling out the over and underestimates of the model. The model may be improved with further effort to correctly take into account relaxation in the transfer line and make polarization inference a more attractive method.

\section{Conclusion}
We have performed several measurements towards the understanding of the dynamics of polarized atoms in transit through depolarizing magnetic gradients. Despite passing through a region of considerable gradients, net polarization was maintained in gas transport via diffusion at 1\,torr into a second cell. 

The dual exponential decay nature of the relaxation of the gas in the two cell system has been approximated by a simple differential model, which allows a limited inference of the polarization in one cell using measurements in the other. Inaccuracy in the model limits the usefulness of this technique, although combined with measurements of relaxation under different initial measurement cell conditions, the predictive power of the model improves. This may be useful in applications where the polarization in the second cell cannot be directly measured.

These measurements provide a basis for coming tests of polarized gas transfer at Brookhaven National Laboratory. Transfer into the 5\,T magnetic field of the EBIS solenoid will first be tested via diffusion at 1 torr, which will lead to tests of gas transfer from 1 torr to $10^{-7}$ torr. Final tests of ionization and extraction of polarized $^3$He from EBIS will follow successful tests of polarized gas transfer in the EBIS stray field.

\subsection*{Acknowledgements}
We gratefully acknowledge Thomas Gentile of NIST for his guidance in MEOP techniques. This research was funded by the program for R\&D for Next Generation Nuclear Physics Accelerator Facilities of the DOE Office of Nuclear Physics.  C.S.E.~acknowledges the support of a DOE NNSA Stewardship Science Graduate Fellowship provided under grant number DE-NA0002135.

\bibliography{pol_trans}
\bibliographystyle{elsarticle-num}

\end{document}